\documentclass[letterpaper]{article} 
\usepackage{algorithm}
\usepackage{algorithmic}
\usepackage[]{aaai25}  
\usepackage{times}  
\usepackage{helvet}  
\usepackage{courier}  
\usepackage[hyphens]{url}  
\usepackage{graphicx} 
\usepackage{natbib}  
\usepackage{caption} 
\usepackage{algorithm}
\usepackage{algorithmic}
\usepackage{algorithm}
\usepackage{algorithmic}
\usepackage{amsmath}
\newtheorem{Definition}{Definition}
\usepackage{subcaption}
\usepackage{multirow}
\usepackage{amsfonts}
\usepackage{mathrsfs}
\usepackage{arydshln}
\usepackage{xcolor}

\usepackage{newfloat}
\usepackage{listings}
\DeclareCaptionStyle{ruled}{labelfont=normalfont,labelsep=colon,strut=off} 
\floatstyle{ruled}
\newfloat{listing}{tb}{lst}{}
\floatname{listing}{Listing}
\title{Learning Complex Heterogeneous Multimodal Fake News via Social Latent Network Inference}
\author{
    Mingxin Li, Yuchen Zhang, Haowei Xu, Xianghua Li\footnote{corresponding author}, Chao Gao, Zhen Wang
}
\affiliations{
    Northwestern Polytechnical University\\

    \{mingxinli, yc20011204, hwxu\}@mail.nwpu.edu.cn, \{li\_xianghua, cgao, w-zhen\}@nwpu.edu.cn
}

\usepackage{bibentry}

\begin{document}

\maketitle

\begin{abstract}

With the diversification of online social platforms, news dissemination has become increasingly complex, heterogeneous, and multimodal, making the fake news detection task more challenging and crucial.
Previous works mainly focus on obtaining social relationships of news via retweets, limiting the accurate detection when real cascades are inaccessible. Given the proven assessment of the spreading influence of events, this paper proposes a method called \textbf{HML} (Complex \textbf{H}eterogeneous \textbf{M}ultimodal Fake News Detection method via \textbf{L}atent Network Inference). Specifically, an improved social latent network inference strategy is designed to estimate the maximum likelihood of news influences under the same event. Meanwhile, a novel heterogeneous graph is built based on social attributes for multimodal news under different events. Further, to better aggregate the relationships among heterogeneous multimodal features, this paper proposes a self-supervised-based multimodal content learning strategy, to enhance, align, fuse and compare heterogeneous modal contents. Based above, a personalized heterogeneous graph representation learning is designed to classify fake news. Extensive experiments demonstrate that the proposed method outperforms the SOTA in real social media news datasets.

\end{abstract}

\section{Introduction}

Nowadays, \textbf{complex heterogeneous multimodal news} (e.g. short videos) has spread and proliferated rapidly due to their sharing nature on social media platforms~\citep{gao2023echo, zheng2023research}. However, the increasing sophistication of forgery techniques has resulted in a large number of multimodal fake news flooding the Internet. These news may involve picture synthesis, audio and video editing, and the application of artificial intelligence technology, making it difficult for audiences to distinguish the real from the fake, and posing a serious challenge to the dissemination of information and social opinion~\citep{bu2023combating, bhargava2023effective}. Therefore, it is an urgent and pressing task to study the detection of complex heterogeneous multimodal fake news.

\begin{figure}[!htb]
    \centering
    \begin{subfigure}{.48\textwidth}
        \centering
        \includegraphics[width=\textwidth]{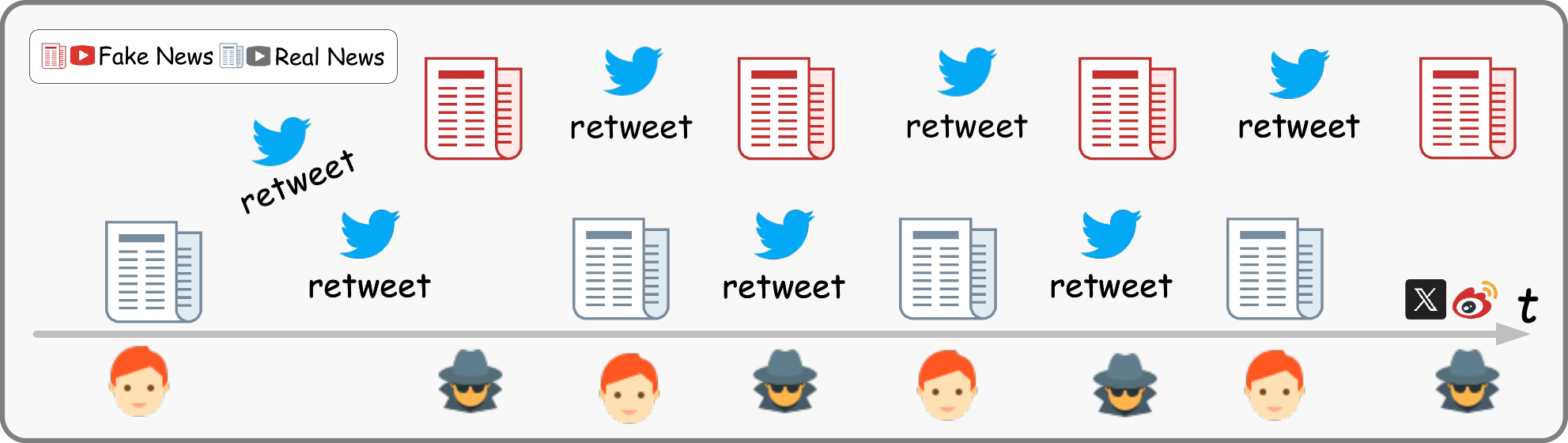}
        \caption{Propagation in Twitter (X) and Weibo.}
    \end{subfigure}%
    \\
    \begin{subfigure}{.48\textwidth}
        \centering
        \includegraphics[width=\textwidth]{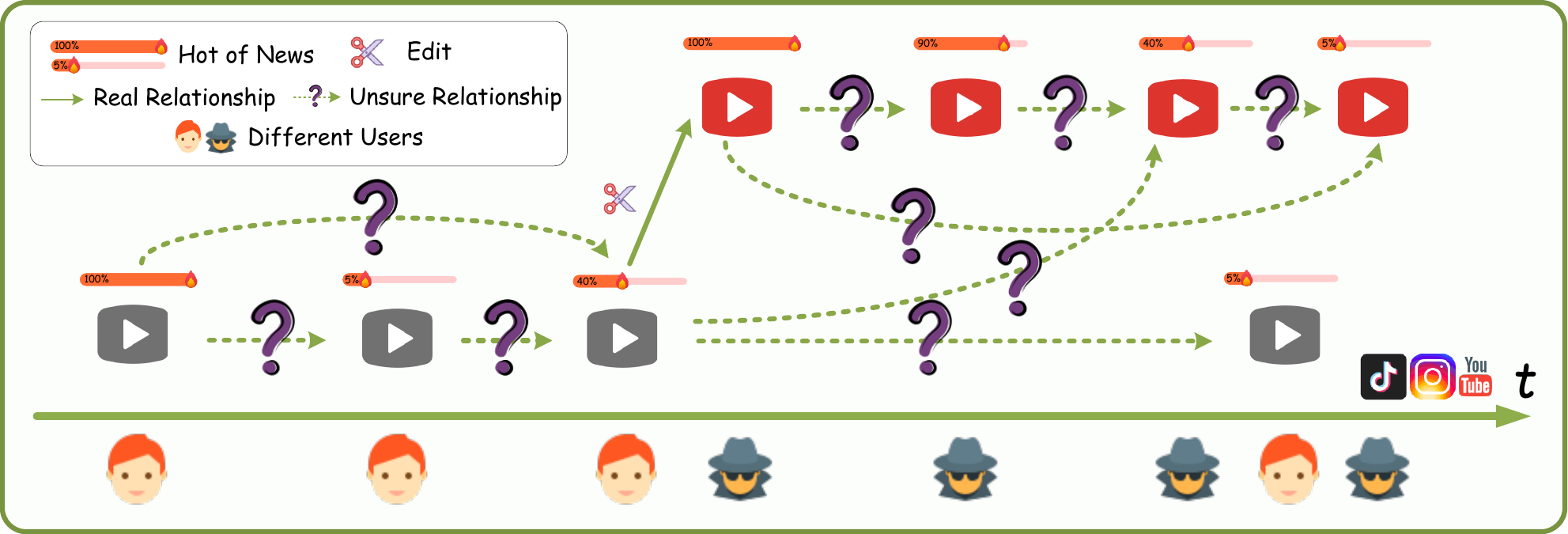}
        \caption{Unsure relationships in TikTok, Instagram and YouTube.}
    \end{subfigure}
    \caption{Differences in news dissemination across social media platforms.} 
    \label{fig:introduction}
\end{figure}

To detect the increasingly complex and varied fake news, some studies have been proposed. To summarize, existing methods are mainly classified into two aspects,  social relationship-based~\cite{shu2020fakenewsnet} and semantic feature-based~\cite{wang2018eann}.  In the first stage, the social relationship is mainly represented between different news through graphs. In recent years, the methods have been used primarily for mining the social cascades through commenting and retweeting mechanism~\cite{cheng2021causal, zhang2023rumor, yin2024gamc, zjy2024kdd}. Graph structure mining is performed through GNN-based methods to accomplish the node classification or graph classification tasks to predict fake news. Further, related works focus on the learning and mining of attribute graphs, fusing graph and modal information for classification tasks~\cite{nguyen2020fang, phan2023fake}. For the other stage, existing methods mainly rely on extracting, fusing~\cite{zhou2020similarity, nan2021mdfend}, and enhancing different modal data~\cite{zhu2022m3fend, qi2021improving}. To avoid simple feature fusion engineering, some researchers perform information integrity through fact checking~\cite{vo2018rise}, reading interest~\cite{wu2023see}, external knowledge~\cite{hu2021compare}, etc. These methods have been confirmed to achieve good results.
 
However, new issues have arisen to be addressed in the increasing online social platforms. As shown in Figure \ref{fig:introduction}(a), when a hot social topic emerges, researchers can get almost complete information cascades on social platforms like Twitter, Weibo, etc. Nevertheless, as shown in Figure \ref{fig:introduction}(b), with the emergence of video-based social media platforms such as TikTok, Instagram, and YouTube, \textbf{it has become increasingly difficult to directly acquire such relationships due to platform mechanics}. In addition, although there are many methods for heterogeneous multimodal fake news detection~\cite{choi2021using, 2021covidtiktok, qi2023fakesv, qi2023two}, all of them are simple splicing of heterogeneous multimodal features, which can only \textbf{primarily rely on pre-trained models combined with attention or transformer, using embeddings for modality alignment in news, which loses the relationship between some modalities}. For deep and fine-grained features such as editing and factual alteration, traditional analytics methods make it difficult to obtain the differences between them and real semantic features. Even powerful AI models can hardly detect deepfake videos effectively. Therefore, \textbf{mining the latent relationships of complex heterogeneous multimodal news and effectively aggregating the features}, becoming a great challenge.

To address the above issues, this paper proposes a novel Complex Heterogeneous Multimodal Fake News Detection method, HML. The method builds news latent cascade relationships through proven effective point time processes and social hot influence and performs latent network inference based on news attributes. Further, for better enhancement, alignment, fusing and comparison of different modal features, this paper proposes a Self-supervised-based Multimodal Content Learning strategy, which can get more effective feature representation. Finally, a personalized heterogeneous graph representation is established to represent the attribute heterogeneous graphs obtained above and to accomplish the fake news detection task. 
Overall, the application of network inference and self-supervised learning can effectively improve the robustness of the model.

For these reasons, the main contributions of this paper are as following three aspects:

\begin{itemize}
    \item \textbf{News Latent Heterogeneous Graph Inference.} Based on the self-excitation of events over time and the influence of event popularity, latent cascade inference is performed. Combining this with news attributes and inferences a latent heterogeneous news graph.
    \item \textbf{Self-supervised-based Multimodal Content Feature Representation.} Content comprehension and augmentation of unimodal content are achieved using self-supervised learning strategy. Additionally, content contrast enhancement of cross-modal helps obtain effective features with robustness.
    \item \textbf{Personalized Representation and Better Performance.} Personalized heterogeneous graph representation of the above features was performed for the Complex Heterogeneous Multimodal Fake News Detection task. Extensive experiments on benchmark datasets show that the proposed method effectively obtains relationships among news and outperforms the SOTA method. 
\end{itemize}

\section{Preliminary}


\begin{Definition}[Attribute Heterogeneous News Graph]
    A attribute heterogeneous news graph is defined as a graph $\mathcal{H} = ( \mathcal{V}, \mathcal{E}, \mathcal{O}_E, \mathbf{A}, \mathcal{X})$ with news nodes $\mathcal{V}$ and multiple types of edges $\mathcal{E}$. $\mathcal{O}_E$  represents the set of object types of edges and $\mathcal{X}$ is the attribute representation of nodes, respectively. In addition, each node is associated with heterogeneous multimodal news.
\end{Definition}

\noindent \textbf{Task 1 (Heterogeneous Latent Network Inference in Social Media Platforms)} 
Let $\mathcal{G} = (\mathcal{V}, \mathcal{E}, \mathcal{O}_E, \mathbf{A})$ represents a latent network where $\mathcal{V}$ is the known set of nodes with $|\mathcal{V}| = n$, $\mathcal{E}$ is the unknown set of edges, and $\mathbf{A} \in \mathbb{R}^{n \times n}$ is the unknown adjacency parameter matrix. Given a set of news attribute relationships $\mathcal{R}$ and latent cascades $\mathcal{C}$, the goal of news latent network inference is to estimate the adjacency parameter matrix $\mathbf{A}$ using $\mathcal{C}$ and $\mathcal{R}$, thereby inferring the underlying edge set $\mathcal{E}$ of the network $\mathcal{G}$.

\noindent \textbf{Task 2 (Complex Heterogeneous Multimodal Fake News Detection)} The dataset is defined as $V=\{v_1, v_2, \ldots, v_n\}$, where each $v_i$ is a news instance. Each instance $\mathbf{x}_i$ has at least 3 complex heterogeneous modalities, such as text, video frame, image, audio et al. Besides, $\mathcal{X}=\{\mathbf{x}_1, \mathbf{x}_2, \ldots, \mathbf{x}_n\}$ are the feature representations of $V$, We aim to learn a self-supervised function, $g$, defined as:
$$
g: V \rightarrow Y,
$$
where $V$ represents news instances with their latent relationship processes and $Y \in \{Fake, Real\}$ denotes either fake or real news.  

\section{Methodology}

\begin{figure}
    \centering
    \includegraphics[width=0.9\linewidth]{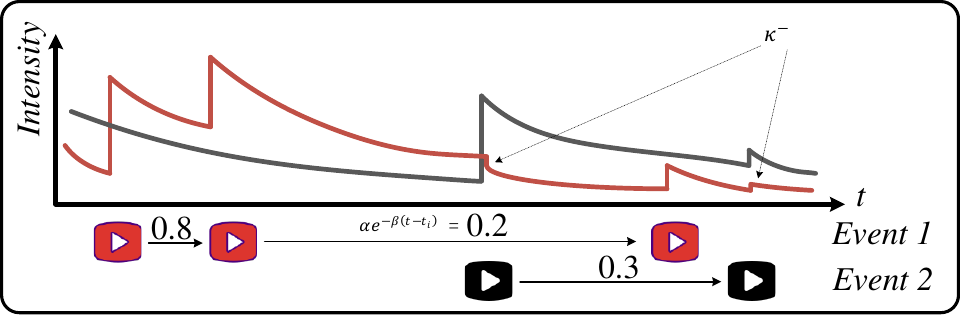}
    \caption{Illustrate of proposed event-based cascade influence. The red line represents the influence intensity of Event 1 over time $t$, which is affected by Event 2 (gray line).}
    \label{fig:influence}
\end{figure}

\begin{figure*}
    \centering
    \includegraphics[width=0.9\linewidth]{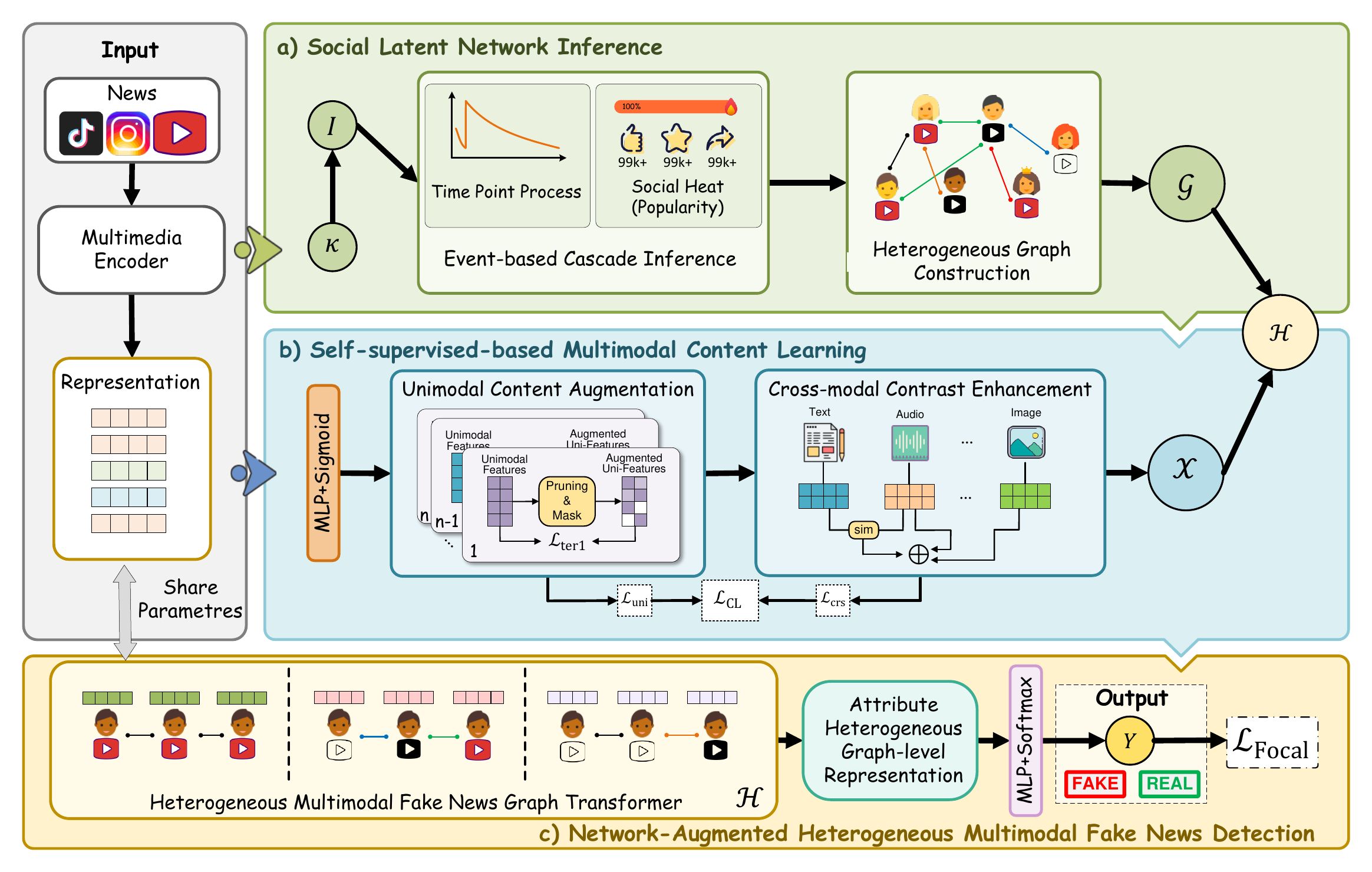}
    \caption{Framework of HML. The whole framework is divided into three modules.}
    \label{fig:framework}
\end{figure*}

This section will introduce the proposed method HML in detail, the illustration of the method is shown in Figure \ref{fig:framework}.

\subsection{Social Latent Network Inference (Stage 1)}
In the stages of news dissemination and evolution, there exist latent cascade paths that change over time, and different news items evolving at different times and states influence each other. Through these relationships, the process of news dissemination and evolution can be effectively established. Therefore, this section innovatively proposes a social latent network inference strategy, which builds a latent social network in the complete absence of real social relationships.

\subsubsection{Event-based Cascade Inference via Hawkes Process}

When a hot social topic emerges, a large number of similar news appear on social network platforms, which are highly correlated and we define them as the same event. The influence intensity of news about the same event in the social network changes over time. When an event suddenly occurs, it often has the strongest influence, but the influence intensity gradually decreases over time. Additionally, the influence can undergo nonlinear changes due to the impact of news from the same event as well as news from different events\cite{matsubara2017nonlinear}. The framework is shown in Figure \ref{fig:framework}. This paper represents the influence of news over time by improved Hawkes process as formula (\ref{eq:hawkes}).

\begin{equation}
\small
\label{eq:hawkes}
    \lambda^*(t) = \lambda_0 + \kappa(t) ,
\end{equation}

\noindent where $\lambda^*(t)$ is the influence intensity function at time t, $\lambda_0$ is the baseline intensity, which is a constant, $\kappa(t)$ represents the influence of past news on the current intensity.The equation of $\kappa(t)$ is specifically expressed as follows:

\begin{equation}
\small
\kappa(t) =
\begin{cases}
\kappa^+ = \sum\limits_{t_i<t} \alpha e^{-\beta(t-t_i)}, &  \text{same event} \\ 
\kappa^- = \sum\limits_{t_i<t} \gamma e^{-\beta(t-t_i)} \cdot \text{prop}(\cdot), &  \text{diff event}
\end{cases}
\end{equation}

\noindent where $\kappa^+$ represents the positive correlated impact of news within the same event due to self-excitation, and $\kappa^-$ represents the impact of news from different events on the current event's news. $\text{prop}(\cdot)$ indicates the correlation between news items, with values ranging from $[-1, 1]$. When a highly popular news item with poor correlation arises, it negatively impacts news from other events. Therefore, the original correlation is mapped using the equation $\text{prop}(\cdot) = \tanh(2 \cdot \text{sim}(\cdot, \cdot) - 1)$. $\alpha$, $\beta$, and $\gamma$ are statistical parameters. The explanation of how to calculate \text{prop}(s) as follows:

During normal dissemination, older news tends to be forgotten more quickly. Consequently, the likelihood of people associating with related news decreases. To simulate this characteristic, this paper introduces Gaussian Noise based on news dissemination, described by the following equation\footnote{Distinct from stage 2, features are not fine-tuned.}:
\begin{equation}
\small
\begin{aligned}
q\left(\mathbf{x}^1, ..., \mathbf{x}^T | \mathbf{x}^0\right) &= \prod_{t=1}^T q\left(\mathbf{x}^t | \mathbf{x}^{t-1}\right) \\
q(\mathbf{x}^t | \mathbf{x}^{t-1}) &= \mathcal{N}(\mathbf{x}^t; \sqrt{\eta_t} \mathbf{x}^{t-1}, (1 - \eta_t) \mathbf{I})
\end{aligned}
,
\end{equation}

\noindent where $T$ is the number of steps to add noise, $\eta$ is a learning variation of time process. This noising process can be defined as a Markov process. To simplify the computation, it can be transformed into the following.

\begin{equation}
\small
\begin{aligned}
q\left(\mathbf{x}^t \mid \mathbf{x}^0\right) & =\mathcal{N}\left(\mathbf{x}^t ; \sqrt{\bar{\delta}_t} \mathbf{x}^0,\left(1-\bar{\delta}_t\right) \mathbf{I}\right) \\
\mathbf{x}^t & =\sqrt{\bar{\delta}_t} \mathbf{x}^0+\sqrt{1-\bar{\delta}_t} \boldsymbol{\epsilon}
\end{aligned}, 
\end{equation}

\noindent $\text { where } \delta_t=1-\eta_t, \bar{\delta}_t=\prod_{s=1}^t \delta_s \text { and } \boldsymbol{\epsilon} \sim \mathcal{N}(\mathbf{0}, \mathbf{I}) \text {. }$

After applying the noising process to news from other events, the similarity $\text{sim}(\cdot, \cdot)$ is calculated using the Minkowski distance as follows:
\begin{equation}
\small
\label{eq:Minkowski}
    \text{prop}(\mathbf{x}_j) = \text{sim}(\mathbf{x}_i^t, \mathbf{x}_j) = (\sum_{k=1}^D |\mathbf{x}_{i,k}^t - \mathbf{x}_{j,k}|^D)^{1/D} , 
\end{equation}

\noindent where $\mathbf{x}_i, \mathbf{x}_j$ are attributes of different nodes, $k$ is an iterator, and $D$ is the feature dimension.

To estimate the statistics parameters, the log likelihood function of the proposed improved news Hawkes process is as follows:
\begin{equation}
\small
    \mathscr{L} = \log \prod_{i=1}^N \lambda^*(t_i)\exp{\left( -\int_0^T\lambda^*(t)dt\right)}  ,
\end{equation}
\noindent where the approximation simplifies the calculation by approximating the integral with a summation over discrete time points.

To calculate the influence of the $i$-th news on the current news at time $t$, we have the following equation:
\begin{equation}
\small
    I(t)_i = \alpha e^{-\beta(t-t_i)} + \kappa^-(t) ,
\end{equation}

\noindent where $\alpha e^{-\beta(t-t_i)}$ represents the influence of the \(i\)-th news on the current news' popularity within the same event, and $\kappa^-(t)$ is the influence of other news on the current news.

After calculating the influence \(I(t)\) of news within the same event, we estimate the adjacency parameter matrix $\mathbf{A}_e$ for the event. At this point, $\mathbf{A}_{e}$ is an \([0,1]^{r \times r}\) matrix. $r$ is the news number of the event (different events have different $r$). This transformed matrix can serve as a prior condition for establishing the heterogeneous news relationship graph described in the next section.

\subsubsection{Heterogeneous Graph Construction}

Unlike widely used heterogeneous graphs, the heterogeneous graph provided in this paper does not contain nodes of different attributes such as paper-author, paper-author-venue, or user-item, all nodes are news items. When news is published on social media platforms such as TikTok or YouTube, a large amount of information about the news and its authors emerges. This information possesses very distinct statistical and sociological characteristics. \citealp{hou2024dag} have statistically proven this point. For example, in the process of news dissemination, compared to real news, fake news often lacks user authentication and exhibits extremely severe differences in title characteristics. However, these features are often not effectively utilized in the complex heterogeneous multimodal fake news detection task, where only simple features are used for feature extraction. Therefore, classifying attributes is key to constructing the proposed heterogeneous graph.

\begin{figure}
    \centering
    \includegraphics[width=1.05\linewidth]{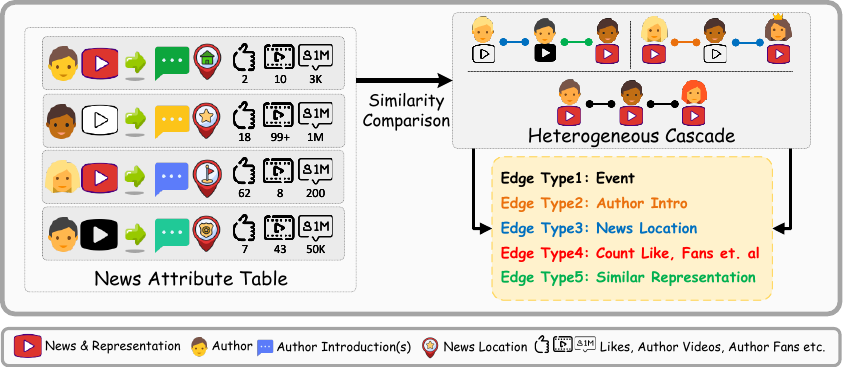}
    \caption{An example of the proposed Heterogeneous Graph Construction strategy in dataset FakeSV.}
    \label{fig:heterougeneous}
\end{figure}

Inspired by \citealp{qi2023fakesv} and \citealp{hou2024propagation}, this paper constructs news using the top few most significant statistical features with the largest differences. As shown in Figure \ref{fig:heterougeneous}, these are the attribute relationships that can be constructed in the FakeSV dataset. Additionally, Some attribute relationships are composed of similarities. However, since this paper assumes no prior knowledge of the real network, a threshold \(\rho\) is designed to map the \([0,1]^{n \times n}\) matrix to a \(\{0,1\}^{n \times n}\) matrix. We have $f(v_i, v_j) =$
$
\begin{cases}
1 &  \text{if}\text{  } \text{sim} \geq \rho \\
0 & \text{if}\text{  } \text{otherwise}
\end{cases}
$
. Among them, the matrices composed of different edge types are named $\textbf{A}_i, i \in \mathcal{O}_E$. It should be noted that $\textbf{A}_e$ is a submatrix of $\textbf{A}_1$. And the setting of edges is shown in Figure \ref{fig:heterougeneous}.

\subsection{Self-supervised-based Multimodal Content Learning (Stage 2)}

\subsubsection{Pruning and Mask}

Data imbalance is a very important issue that affects the performance of comparative learning. Nowadays, there have been many works to solve the problem of data imbalance~\cite{jiang2021self, frankle2018lottery}. In this paper, we follow a pruning method for data augmentation on unbalanced data~\cite{frankle2018lottery}.

To perform effective data augmentation, some modal features of the news are masked to reconstruct their initial features, resulting in new effective representations. The masking method for features can be described as  $\hat{\mathbf{x}_i}$ = 
$
\begin{cases}
\mathbf{x}_{[MASK]} &  \text{if}\text{  }  v_i \in V^m \\
\mathbf{x}_i & \text{if}\text{  }  v_i \notin V^m
\end{cases}
$
,$V^m$ is the news mask list.

\subsubsection{Unimodal Content Augmentation}

This section will be devoted to the unimodal content augmentation strategy used. Unimodal content learning is mainly applied to the analysis of correlations within similar modalities, where the original modal features are compared with the augmented features to calculate the loss with the following equation:
\begin{equation}
    \mathcal{L}_{\text {uni}}=-\log \sum_{\mathbf{x}_i \in \mathcal{X}} \frac{\exp \left[\operatorname{sim}\left(\widetilde{z}^i, \widetilde{z}^j_p\right) / \tau_{\text {uni }}\right]}{\sum_{p=1}^{2 n} \mathbb{1}_{[i \neq p]} \exp \left[\operatorname{sim}\left(\widetilde{z}^i, \widetilde{z}^p_p\right) / \tau_{\text {uni }}\right]},
\end{equation}

\noindent where $\widetilde{z}$ represents the processed correlation feature. $\widetilde{z}^i, \widetilde{z}^j, \widetilde{z}^p$ are unpruned data, $\tilde{z}_{p}^j, \tilde{z}_{p}^p$ are augmented data. $\tau_{\text {uni }}$ is the temperature value, which is a hyperparameter.

\subsubsection{Cross-modal Contrast Enhancement}

The strategy is mainly used for the comparison between different modal data. All contrast losses are calculated and summed to get the cross-modal contrast loss, 
\begin{equation}
    \mathcal{L}_{\text {crs }}=-\sum_{i}\log \sum_{\mathbf{x}_i \in \mathcal{X}} \frac{\exp \left[\operatorname{sim}\left(\widetilde{z}^i, \widetilde{z}^j\right) / \tau_{\text {crs }}\right]}{\sum_{p=1}^{2 n} \mathbb{1}_{[i \neq p]} \exp \left[\operatorname{sim}\left(\widetilde{z}^i, \widetilde{z}^p\right) / \tau_{\text {crs }}\right]}.
\end{equation}

\subsubsection{Multimodal Loss Integration}

To integrate the above strategies, a hyperparameter $\lambda$ is set for joint loss to use for tuning the network.

\begin{equation}
\small
\mathcal{L}_{CL} = \lambda \mathcal{L}_{\text {uni }}+(1-\lambda) \mathcal{L}_{\text {crs}} + ||\mathbf{\Theta}||_2 ,
\end{equation}
\noindent where $||\mathbf{\Theta}||_2$ represents the L2 normalization. The parameters of above are updated through the above process to prepare for the next step.

\subsection{Network-Augmented Heterogeneous Multimodal Fake News Detection Task}

According to the model after parameter update, new latent graph and attributes are obtained for the personalized node classification task.

\subsubsection{Edge Types Combination}

The previous heterogeneous graph construction has generated edges of different types. These edges obviously have different contributions to the graph. Thus, it needs to be represented differently. We employ an attention mechanism to solve the difficulties. The aggregate method is formulated as:

\begin{equation}
\small
    \mathbf{A}=\sum_{i \in \mathcal{O}_E} \omega^{v \leftrightarrow v} \mathbf{A}_{i} ,
\end{equation}

\noindent where $\mathbf{A} \in \mathbb{R}^{n \times n}$ is the final heterogeneous adjacency matrix, $\omega^{v \leftrightarrow v}$ is the importance of different types of edges, which is learnable in attention mechanism.

Finally, to get the effective representation, this paper explains a useful mechanism of graph representation, graph transformer encoder~\cite{zjy2024kdd}. Unlike other methods, this approach combines modal features and graph structures to obtain an effectively integrated feature representation. The equation is designed below:

\begin{equation}
\small
    \operatorname{Attn}^h(\mathbf{Q}, \mathbf{K}, \mathbf{A})=\operatorname{softmax}\left(\mathcal{M}\left(\frac{\mathbf{Q}^h \mathbf{K}^{h T}}{\sqrt{d_H}}, \mathbf{A}\right)\right) ,
\end{equation}

\noindent where $\mathcal{M}$ is masking function, defined as: $\mathcal{M}(u,v)=u+\zeta v$, $\zeta$ is a sufficiently large value. 

\subsubsection{Classification}

To improve the robustness of the model, a focal loss~\cite{lin2017focal} is introduced in this paper, as shown in equation (\ref{eq:focalloss}). By designing two hyperparameters, the nodes that are difficult to classify are centralized.
\begin{equation}
\label{eq:focalloss}
\small
    \mathcal{L}_{\text {focal }}=-\frac{1}{\left|\mathcal{V}_l\right|} \sum_{i \in \mathcal{V}_l} \sum_{c=0}^C \phi_c y_{i c}\left(1-\hat{y}_{i c}\right)^\psi \log \left(\hat{y}_{i c}\right) , 
\end{equation}

\noindent where $\mathcal{V}_l$ is the node containing the label, $C$ is the category, and $\phi_i, \psi \in [0,1]$, is a tunable hyperparameter. $y_{ic}$ is the actual label value and $\hat{y}_{ic}$ is the predicted label value.

\section{Experiments}

In this section, we conduct some extensive experiments to evaluate the proposed framework, HML. 

\subsection{Benchmark Datasets}

\begin{table*}
\small
    \centering
    \caption{Experimental results of different methods. The experiments are conducted to compare unimodal, multimodal, and large language models.  Besides, the last line shows the enhancement of the proposed method to baseline methods. \underline{Underline} denotes the second-best metric, while \textbf{bold} denotes the best metric. (Acc.: accuracy, Prec.: precision, Rec.: recall)} 
    \begin{tabular}{cccccccccc}
        \hline \hline 

        & \multirow{2}{*}{\textbf{Method}} & \multicolumn{4}{c}{\textbf{FakeSV}} & \multicolumn{4}{c}{\textbf{FVC}}\\
        &   & \textbf{Acc.} & \textbf{Prec.} & \textbf{Rec.} & \textbf{F1}  & \textbf{Acc.} & \textbf{Prec.} & \textbf{Rec.} & \textbf{F1}\\
        \hline
        \multirow{4}{*}{Unimodal} & Text (BERT)                & 77.06  & 77.16  & 77.07 & 77.04 & 61.70  & 61.81  & 61.72 & 61.76\\
        &Audio (VGGish)              & 66.72 & 67.05 & 66.70 & 66.53  & 58.44  & 58.48  & 58.63 & 58.61\\
        &Image (VGG19)               & 69.66 & 69.78 & 69.78 & 69.59 & 65.79  & 65.49  & 66.08 & 65.81\\
        &Video (C3D)                 & 69.59 & 70.07 & 69.56 & 69.38  & 71.81  & 71.89  & 71.85 & 71.72\\
        \hdashline[1pt/2pt]
        \multirow{6}{*}{Multimodal} & Serrano et al.   & 68.71 & 70.74 & 68.73 & 67.91  & 66.87  & 67.15  & 66.34 & 66.74\\
        &FANVM         & 76.37 & 75.39  & 73.71  & 74.18  & 85.81  & 85.20  & 85.44 & 85.35\\
        &SV-FEND                    & 78.88  & 79.41 & 78.89 & 78.79  & 84.71  & 84.25  & 86.53 & 85.37\\
        &MMVD                    & 82.64  & 82.63 & 82.73 & 82.63  & \underline{89.28}  & \underline{90.27}  & \textbf{90.36} & \underline{90.46}\\
        &NEED                    & 84.62  & 84.81 & \underline{84.64} & 84.61  & -  & -  & - & -\\
        &FakingRecipe            & \underline{85.35}  & \underline{85.84} & 84.29 & \underline{84.83}  & -  & -  & - & -\\
        \hdashline[1pt/2pt]
        \multirow{3}{*}{LLM} & Doubao        & 70.25     & 72.60  & 70.24  & 69.45  & 45.89  & 44.40  & 45.89 & 44.82\\
        &GPT-4o               & 73.07  & 73.31  & 73.06 & 72.99  & 50.72  & 49.94 & 50.72 & 50.17\\
        &ARG &  78.32 & 78.60 & 78.36 & 78.28  & 84.44  & 85.19  & 83.93 & 84.16 \\
        \hdashline[1pt/2pt]
        & \textbf{HML(Ours)}  & $\textbf{89.14}_\text{(+3.79\%)}$ & \textbf{89.36 } & \textbf{89.22 } & $\textbf{89.22 }_\text{(+4.39\%)}$ & $\textbf{91.02}_\text{(+1.74\%)}$ & \textbf{91.68 } & \underline{90.32}  & $\textbf{90.58}_\text{(+0.12\%)}$\\
        \hline \hline
    \end{tabular}
    \label{tab:experiment_not_class}
\end{table*}

\begin{table}
\small
    \centering
    \caption{Plugin Experiments on Twitter and Weibo. \\ NI: Network Inference}
    \begin{tabular}{cccccc}
        \hline \hline
        \multirow{2}{*}{\textbf{Method}}& \multicolumn{2}{c}{\textbf{Twitter}} & \multicolumn{2}{c}{\textbf{Weibo}} \\
          & \textbf{Acc.} & \textbf{F1} & \textbf{Acc.} & \textbf{F1} \\

        \hline
        EANN     & 64.86 & 63.92  & 79.56 & 80.03\\
         \textit{+NI}  & \textbf{74.72}    & \textbf{74.63}  & \textbf{84.89} & \textbf{84.38}    \\
           & (+9.86\%)    & (+10.71\%)  & (+5.33\%) & (+4.35\%) \\
        \hdashline[1pt/2pt]
        MCAN    & 74.55 & 74.85  & 89.96 & 89.33\\
         \textit{+NI}   & \textbf{77.45} & \textbf{77.92}  & \textbf{90.27} & \textbf{90.60}     \\
            & (+2.90\%)    & (+3.07\%)  & (+0.31\%) & (+1.27\%) \\
        \hdashline[1pt/2pt]
        CAFE   & 80.62 & 80.38  & 84.13 & 83.77\\
          \textit{+NI}  & \textbf{83.40} & \textbf{82.94}  & \textbf{85.91} & \textbf{85.68}    \\
            & (+2.78\%)    & (+2.56\%)  & (+1.78\%) & (+1.91\%) \\
        \hline \hline
    \end{tabular}
    
    \label{tab:plugin}
\end{table}

This paper applies two complex heterogeneous multimodal fake news detection datasets, FakeSV~\cite{qi2023fakesv} and FVC~\cite{papadopoulou2019corpus}. Additionally, to prove the social latent graph inference method can be widely applied to multimodal fake news datasets, this paper additionally uses Twitter~\cite{boididou2018detection} and Weibo~\cite{wang2018eann} datasets to evaluate the inference task as a plugin.
\subsection{Baselines}

In order to prove the superiority of the method, some of the more advanced algorithms are compared as follows:
\subsubsection{Unimodal}
Traditional analytical methods mainly explore the expressive unimodal features. This paper mainly use \textbf{BERT}~\cite{devlin2018bert}, \textbf{VGGish}~\cite{hershey2017vggish}, \textbf{VGG19}~\cite{simonyan2014vgg19}, and \textbf{C3D}~\cite{ji2012c3d} to analyze the characteristics of their respective modalities.

\subsubsection{Multi-modal}

Existing multimodal methods mainly focus on cross-modal methods , such as \textbf{\citealp{serrano2020nlp}}, \textbf{FANVM}~\cite{choi2021using}, \textbf{SV-FEND}~\cite{qi2023fakesv}, \textbf{MMVD}~\cite{mmvd}, \textbf{NEED}~\cite{qi2023two}, and \textbf{FakingRecipe}~\cite{bu2024fakingrecipe}. However, fewer analyses have both semantic features and social features. This paper will compare the aforementioned work with the proposed method HML. 
\subsubsection{Large Language Model}

To explore the application ability of LLMs for the task of fake news detection, this paper designs the prompt method and conducts experiments through the APIs of \textbf{Doubao}\footnote{\url{https://www.doubao.com}} and \textbf{GPT-4o}\footnote{\url{https://chatgpt.com}}. Additionally, a baseline model, \textbf{ARG}~\cite{hu2024badarg}, for detecting fake news in LLM has also been evaluated. 

\subsection{Experimental Setup}

\subsubsection{Metrics and Parameters} 

To prevent the model from capturing features of news evolution, this paper divides the dataset 80:20 following the FakeSV benchmark. To validate the effectiveness of the model, this paper uses a computational classification accuracy method to evaluate the effectiveness of the model using Accuracy, Precision, Recall, and F1-score with interval estimation and K-S test. For parameters, this paper uses 4 $\times$ NVIDIA RTX A6000 as GPU. Besides, the method achieves optimal performance with $\lambda=0.5$ and pruning ratio $e=0.6$.

\subsection{Overall Performance}

We have compared our methods to unimodal methods, multimodal methods, and large language model prompt methods respectively, and the comparison results obtained are shown in Table \ref{tab:experiment_not_class}, which are analyzed as follows,

\subsubsection{Performance}
After the above experiments, all the metrics of our method reach more than 89\% in the FakeSV dataset and 90\% in the FVC dataset, which is an improvement of $\text{0.12\%} \sim \text{4.39\%}$ compared with the previous best method in both datasets. After comparing with the unimodal method and LLM method, it is found that the proposed method substantially improves the relevant metrics. 
\subsubsection{Analysis} 
    For unimodal classification methods, only a very small portion of the news information can be learned. For prior multimodal methods, the semantic understanding methods for news are mainly unimodal and cross-modal simple feature extraction fusion and classification, which cannot effectively recognize the fine-grained differences between news. Our approach HML both alignes, fuses multimodal features, and mines latent networks between news. In addition, contrastive learning is used to amplify the differences between different samples. Thus, our method outperforms the state-of-the-art method.
\subsubsection{Compare with LLM} According to the experimental results, it can be seen that LLM performs poorly in the multimodal fake news detection task, mainly because the prompt task mainly calls the original features of the model, and it cannot learn from the newly generated news. Therefore, how to adjust LLM to adapt the multimodal fake news detection task may be a topic that can be investigated in the next step.

\subsubsection{Extensive Applications} To demonstrate the effectiveness of the method on a broad range of datasets, this section evaluates it on traditional image-text fake news detection datasets. The network inference method proposed in this paper is used as a plugin and combined with existing classical methods to create a improved fake news detection method. As is shown in Table \ref{tab:plugin}, evaluations are conducted using the EANN~\cite{wang2018eann}, MCAN~\cite{wu2021multimodal}, and CAFE~\cite{chen2022cross} methods, with improvements ranging from 0.3\% to 10\% relative to the baseline. Combined with the ablation experiment results in the next section, our analysis suggests that the proposed method for establishing the latent network can effectively capture features. On the same dataset, since the same latent network structure is obtained, the improvement in metrics depends on the feature extraction capability of the baseline method.

\subsection{Ablation Study}

In order to prove the effectiveness of the proposed strategies, this section investigates the ablation experiments. Experiment results of FakeSV see Table \ref{tab:Ablation_results}.
\subsubsection{Ablation Study about Stage 1}

To ensure that each component in the latent network inference has a positive effect on the whole framework, we designed 3 ablation experiments.

\begin{itemize}
    \item \textbf{w/o Latent Graph (LG)} remove inference the latent (heterogeneous) graph.
    \begin{itemize}
        \item \textbf{w/ LG w/o Event Inference} remove the event based latent inference. 
        \item \textbf{w/ LG w/o Edge Type Construction} remove performing the establishment of heterogeneous graph with edge types. 
    \end{itemize}
\end{itemize}

The experimental results prove the necessity of the latent network inference. Considering only the relationship inherent in the events, the experiment proved that the indicator was reduced by 0.52\%, indicating that there is a latent relationship between different events. For removing the edge-type construction strategy, only the similarity relationship between the news is considered and the event factor is not taken into account. The experiment proves that the indicator is reduced by 2.29\%. However, if the latent network inference method is completely removed, the experimental results drastically drop by 8\%, effectively proving the importance of the proposed strategy.

\begin{table}
\small
    \centering
    \caption{Ablation study on FakeSV.}
    \begin{tabular}{lccccc}
        \hline \hline
        \textbf{Method}  & \textbf{Acc.} & \textbf{Prec.} & \textbf{Rec.} & \textbf{F1} \\
        \hline
        \textbf{HML(Ours)}  & \textbf{89.14} & \textbf{89.36} & \textbf{89.22} & \textbf{89.22} \\
        \hdashline[1pt/2pt]
        w/o Stage1     & 81.68 & 82.21 & 81.31 & 81.44 \\
        - Event Inference   & 86.84     & 86.91 & 86.87 & 86.93  \\
        - Edge Type Cons.    & 88.70     & 89.06 & 88.98 & 88.70  \\
        \hdashline[1pt/2pt]
        w/o Stage 2     & 78.57 & 78.30 & 77.89 & 78.20 \\
        - Unimodal  Augmentation   & 88.80     & 88.95 & 89.00 & 88.79  \\
        - Cross-modal Enhancement    & 88.40     & 88.41 & 88.51 & 88.39 \\
        \hline \hline
    \end{tabular}
    
    \label{tab:Ablation_results}
\end{table}

\subsubsection{Ablation Study about Stage 2}
To justify the Self-supervised-based Multimodal Content Learning, this section conducts ablation experiments on the proposed strategy by eliminating the Unimodal Content Augmentation and Cross-modal Contrast Enhancement, respectively.

\begin{itemize}
    \item \textbf{w/o Multimodal Content Learning (MCL)} remove all the content learning strategy.
    \begin{itemize}
        \item \textbf{w/ MCL w/o Unimodal Augmentation} remove the unimodal augmentation and retain only the cross-modal enhancement strategy.
        \item \textbf{w/ MCL w/o Cross-modal Enhancement} remove the cross-modal enhancement strategy and retain only the unimodal augmentation.
    \end{itemize}

\end{itemize}

Two content learning strategies are removed separately and it can be seen that there is a certain decrease of about 10\% in metrics after removal. 
This indicates that relying solely on the latent network inference is insufficient for content learning to effectively capture the important feature relationships in news.
Furthermore, though the metrics for removing content learning separately do not show more significant improvement compared to stage 1, it still achieves the purpose of proposing this strategy.
For instance, for unlabeled data, the proposed strategy can also enhance the performance through augmentation and comparison.

\begin{figure}[!htb]
    \centering
    \begin{subfigure}{.15\textwidth}
        \centering
        \includegraphics[width=\textwidth]{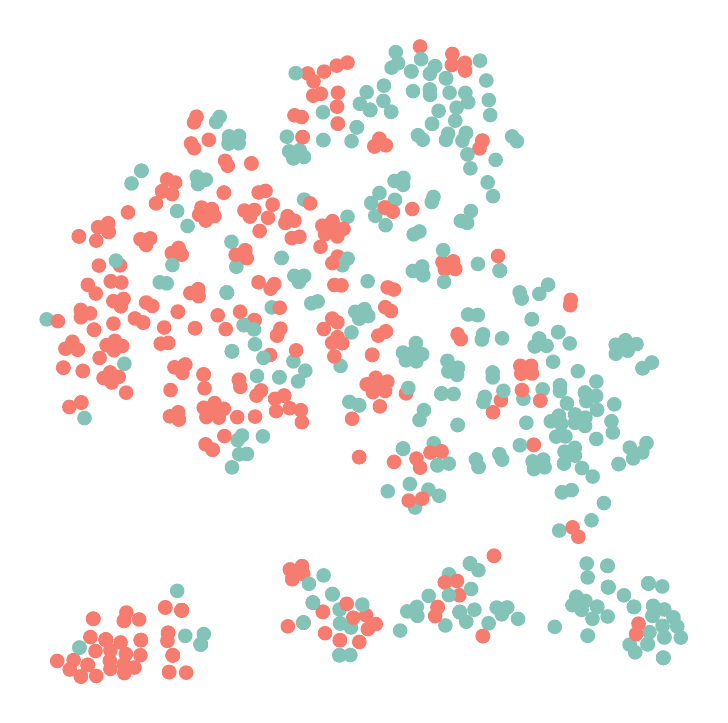}
        \caption{BERT}
    \end{subfigure}%
    \begin{subfigure}{.15\textwidth}
        \centering
        \includegraphics[width=\textwidth]{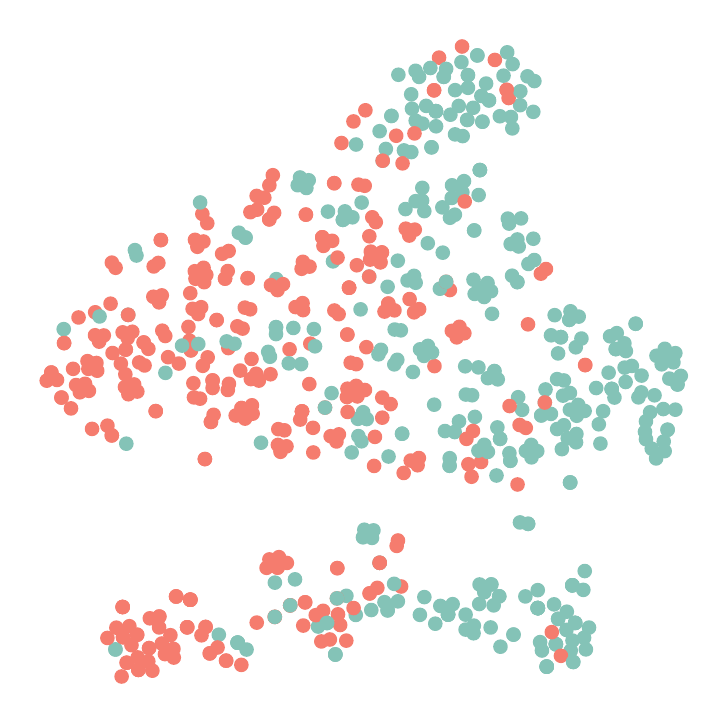}
        \caption{SV-FEND}
    \end{subfigure}
    \begin{subfigure}{.15\textwidth}
        \centering
        \includegraphics[width=\textwidth]{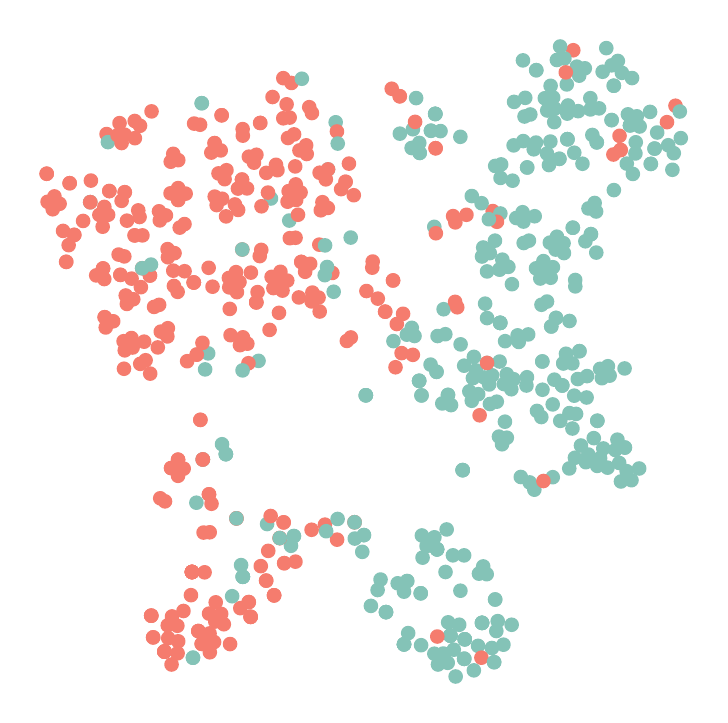}
        \caption{HML}
    \end{subfigure}
    \caption{T-SNE Visualization of FakeSV dataset.} 
    \label{fig:tsne}
\end{figure}

\begin{figure}
    \centering
    \includegraphics[width=\linewidth]{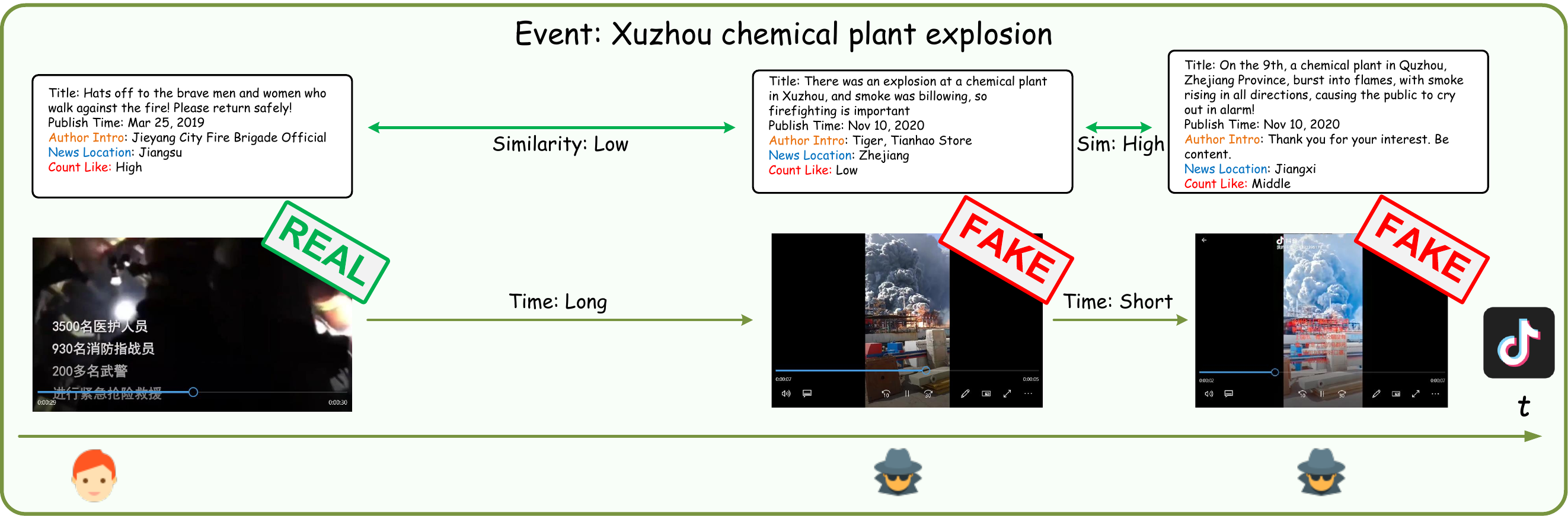}
    \caption{Case study of proposed method HML in Douyin social media platform.}
    \label{fig:case_study}
\end{figure}
\subsection{Visualization and Case Study}
%
\subsubsection{Visualization} This section presents a dimensionality reduction visualization of the classification features for fake news detection. Figure \ref{fig:tsne} shows the feature distributions obtained by different methods on the test set of FakeSV dataset, which contains a total of 717 news articles, with 352 being real and 365 being fake. It is clearly evident that the proposed method clusters news contents of the same category more compactly, qualitatively demonstrating the effectiveness of the method.

\subsubsection{Case Study}

This paper uses the explosion incident in Xuzhou, Jiangsu as an example to illustrate this proposed method. As shown in Figure \ref{fig:case_study}, the chemical plant incident attracted widespread attention in a short period, and the related news received a large number of likes and shares. Official media typically release news immediately after an incident occurs, while fake news often does not meet these conditions. For example, the second piece of news was circulated a year after the event was forgotten, with some media editors reusing old news and adding their own commentary. These self-media sources are usually unofficial, with IP addresses inconsistent with the event location, and have poor relevance, making them identifiable as fake news. By reasoning through the latent network rather than the propagation relationship, fake news can be detected more accurately. The third video, similar to the second one, also meets the criteria for fake news, and when the second piece of news is detected, the third can also be identified.

\section{Conclusion}

This paper explores the latent network inference and the implementation of modal enhancement methods for complex heterogeneous multimodal news. 
Specifically, the social latent network is obtained through event-based cascade inference and relation-based heterogeneous graph construction strategies. Additionally, a news modality learning method is designed to enhance news features and facilitate expansion to large-scale datasets. Extensive experimental results show that the proposed method can get effective network representation capabilities, and the fake news detection method achieves state-of-the-art performance on benchmark datasets.

\section{Acknowledgments}
This work was supported in part by the National Natural Science Foundation of China (Nos. 62271411, 62471403, U22A2098, 62261136549, U22B2036), the Fundamental Research Funds for the Central Universities (Nos. G2024WD0151, D5000240309), and the XPLORER PRIZE; Sponsored by the Practice and Innovation Funds for Graduate Students of Northwestern Polytechnical University.

\bibliography{aaai25}

\end{document}